\documentclass[pra,aps,twocolumn,reprint,floatfix,showpacs,10pt]{revtex4-1}
\usepackage{graphicx}
\usepackage{epsfig}
\usepackage{graphics}
\usepackage{bm}
\usepackage{psfrag}
\usepackage{mathtools}
\usepackage[english]{babel}
\usepackage[T1]{fontenc}
\usepackage{color}
\usepackage[dvipsnames]{xcolor}
\usepackage{braket}

\usepackage{umoline}
\usepackage[colorlinks,bookmarks=false,citecolor=blue,linkcolor=magenta,urlcolor=black]{hyperref}
\usepackage[caption=false]{subfig}

\def\urlprefix{}
\def\url#1{}
\def\be{\begin{equation}}
\def\ee{\end{equation}}
\def\bea{\begin{eqnarray}}
\def\eea{\end{eqnarray}}
\def\bi{\begin{itemize}}
\def\ei{\end{itemize}}
\def\bin{\begin{enumerate}}
\def\ein{\end{enumerate}}
\def\la{\langle}
\def\ra{\rangle}

\begin{document}
\title{Discrete disorder models for many-body localization}
\author{Jakub Janarek}
\affiliation{Instytut Fizyki im. Mariana Smoluchowskiego, Uniwersytet Jagiello\'nski, ul. Profesora Stanis\l awa \L ojasiewicza 11, PL-30-348 Krak\'ow, Poland}
\email{janarek.jakub@gmail.com}

\author{Dominique Delande}
\affiliation{Laboratoire Kastler Brossel, UPMC-Sorbonne Universit\'es, CNRS,
ENS-PSL Research University, Coll\`ege de France, 4 Place Jussieu, 75005 Paris, France and\\
Instytut Fizyki im. Mariana Smoluchowskiego, Uniwersytet Jagiello\'nski, ul. Profesora Stanis\l awa \L ojasiewicza 11, PL-30-348 Krak\'ow, Poland}
\email{dominique.delande@lkb.ens.fr}

\author{Jakub Zakrzewski}
\affiliation{Instytut Fizyki im. Mariana Smoluchowskiego, Uniwersytet Jagiello\'nski, ul. Profesora Stanis\l awa \L ojasiewicza 11, PL-30-348 Krak\'ow, Poland and\\
Mark Kac Complex Systems Research Center, Uniwersytet Jagiello\'nski, Krak\'ow, Poland}
\email{jakub.zakrzewski@uj.edu.pl}

\date{\today}
\begin{abstract}
\noindent Using exact diagonalization technique, we investigate the many-body localization  phenomenon in the 1D Heisenberg chain comparing several disorder models. In particular we consider a family of discrete distributions of disorder strengths and compare the results with the standard uniform distribution. Both statistical properties of energy levels and the long time non-ergodic behavior are discussed. The results for different discrete distributions are essentially identical to those obtained for the continuous distribution, provided the disorder strength is rescaled 
by the standard deviation of the random distribution. Only for the binary distribution significant deviations  are observed.
\end{abstract}


\maketitle

\section{Introduction}
The phenomenon of localization caused by disorder was introduced by Anderson in 1958 \cite{Anderson58}. Although the original Anderson's idea was to describe many-body electronic systems in disordered potentials, his model was relevant for noninteracting particles only. Different attempts to understand localization of interacting particles in disordered potentials have been made, see e.g. \cite{Altshuler79a,Altshuler80,Fleishman1980,Shepelyansky94,Plazmon96} just to give few examples. The breakthrough came with the paper of Basko \textit{et al.} \cite{Basko06}. Since then, the phenomenon of many-body localization {(MBL)}, i.e. localization in the presence of interactions, has been extensively studied in many theoretical works (for recent reviews see, e.g. \cite{Huse14,Rahul15,Abanin17,Alet17})
as well as in experiments \cite{Schreiber15,Kondov15,Bordia16,Choi16,Lueschen17a,Lueschen17b}.

A huge interest in this topic follows from the fact that MBL breaks the eigenstate thermalization hypothesis (ETH) \cite{Deutsch91,Srednicki94} which was assumed to be true for generic interacting many-body systems. ETH states that, because of interactions, the local information on the initial state is smeared over the whole system during time evolution. Thus, in a large enough system, local measurements are unable to retrieve this information and memory of the initial state is lost. Systems which are many-body localized have the exceptional character that, even after long time evolution, they preserve some kind of information about the initial state.

The simplest models for studying MBL are one-dimensional chains or lattices where the disorder is introduced via a random on-site potential or interaction. The most common choice is a {random} uniform continuous distribution in a given interval $[-W,W]$, with $W$ referred as the strength of disorder. However, many other distributions can be used. Experimentally, the disorder can be created by a speckle potential \cite{Damski2003}, the use of different atomic species \cite{Gavish05}  -- in which case the disorder distribution is discrete -- or incommensurate laser waves \cite{Roth03b,Damski2003,Schreiber15} where the disorder distribution is not \textit{stricto sensu} random, but rather pseudo{-}random.

In the standard approach, an average over disorder is performed by explicit averaging over a finite number of disorder realizations. Several years ago, it was suggested \cite{Paredes05} that quantum averaging may be used for that purpose. By adding ancillary spins to the system  and properly constructing the dynamics of the composite system, the trace over the ancillary spins results in ``exact'' averaging.  The method has been recently applied \cite{Enss17} to study the  disordered Heisenberg chain in the thermodynamic limit. The ancillary spins enlarge the size of the Hilbert space. This is strongly undesirable for exact diagonalization methods (see e.g. \cite{Pal10,Luitz15,Luitz16}), which can be performed only for small systems. On the other hand, quantum averaging could be advantageous for tensor-network based algorithms, where the cost of numerical calculation increase only slowly with the system size. The method works only for discrete distributions with the number of possible on-site values of 
disorder directly related to the length of the ancillary spin:  spin-1/2 corresponds to 2 possible values, spin-1 to 3, and so on. The larger the spin, the more computationally demanding is the method. It is interesting to see to what extent the MBL properties depend on the type of discrete distribution used and how fast they converge to the continuous limit. This is addressed in the present paper.

For our analysis, we use the  Heisenberg spin-1/2 chain:
\begin{equation}
	\mathcal{H} = J \sum_{i=1}^L \vec{S}_i \cdot \vec{S}_{i+1} + \sum_{i=1}^{L} w_i S_i^z,
\end{equation}
where $w_i$ are randomly drawn from some distribution defined in the interval $[-W,W]$. For a uniform distribution, it is known that the system exhibits
a transition from an ergodic phase at small $W$ to a MBL one at large $W$ \cite{Huse14}. In our work we compare results computed for several discrete distributions  to the uniform one. In particular we use the binary disorder which corresponds to only 2 possible choices, namely $\{-W,W\}$, the ternary disorder with $\{-W,0,W\}$, the quaternary one $\{-W, -W/3 , W/3, W\}$, as well as the quinary disorder with values: $\{-W, -W/2, 0, W/2, W\}$. All distributions are symmetric with respect to $W=0$, have zero mean, but have different variances. All results have been computed by exact diagonalization for the spin chain of length $L=16$  {with periodic boundary conditions (PBC). As discussed below in Section~\ref{sec:binary_puzzle}, the boundary conditions
are mostly irrelevant, especially for large systems, but should nevertheless be investigated thoroughly.}

\section{Level spacing analysis}

 One of the standard techniques to characterize the transition between the ergodic and MBL states considers statistics of energy levels. In the ergodic phase, one expects  the statistical properties of energy levels to follow (for time-reversal invariant models) the predictions of the Gaussian Orthogonal Ensemble (GOE) of random matrices  \cite{bgs84,Haake04}. Traditional studies in the last millennium  -- e.g. in the quantum chaos field -- involved the cumbersome procedure of unfolding the energy levels, i.e. renormalizing them to have a unit mean density. A significantly simpler idea was given in 2007 by Oganesyan and Huse \cite{Oganesyan07}. They introduced a dimensionless measure of correlation between consecutive energy levels. Let $\delta_n$ be the spacing between consecutive energy levels:
    \begin{equation}
        \delta_n = E_{n+1} - E_n \geq 0,
    \end{equation}
    then the ratio of consecutive spacings is defined as:
    \begin{equation}
        r_n = \min\{\delta_n,\delta_{n-1}\} / \max\{\delta_n,\delta_{n-1}\}.
    \end{equation}
This quantity is normalized $r_n \in [0 ,1]$ and dimensionless. For the GOE ensemble, the ratio can be calculated approximately \cite{Atas13} yielding the mean $\langle r\rangle_{GOE} = 0.53$. In the opposite case, when the system is localized, the spectrum is assumed to follow a Poisson distribution, with the mean ratio equal to $\langle r\rangle_{Poisson} = 2\log 2 -1 \approx 0.39$. In the transition regime between these two  phases, one expects that the mean ratio has intermediate values, smoothly interpolating between {the} limiting cases. Such a situation has been indeed observed in a number of studies \cite{Oganesyan07,Pal10,Luitz15,Luitz16,Sierant17,Sierant17b,Sierant17c}.

\begin{figure*}
        \centering
        \subfloat[][binary]{\includegraphics[scale=0.6]{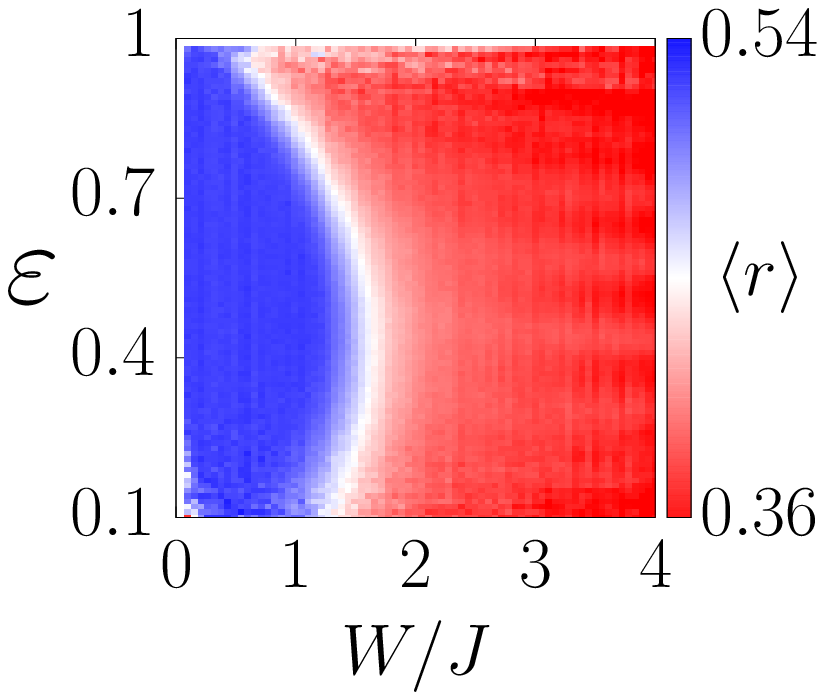}}\quad
        \subfloat[][ternary]{\includegraphics[scale=0.6]{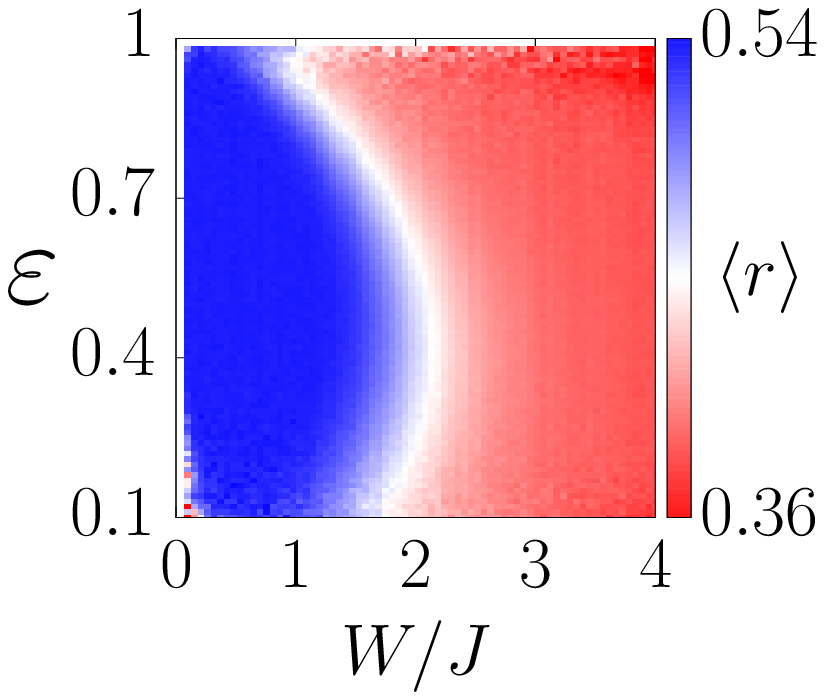}}\quad
        \subfloat[][quaternary]{\includegraphics[scale=0.6]{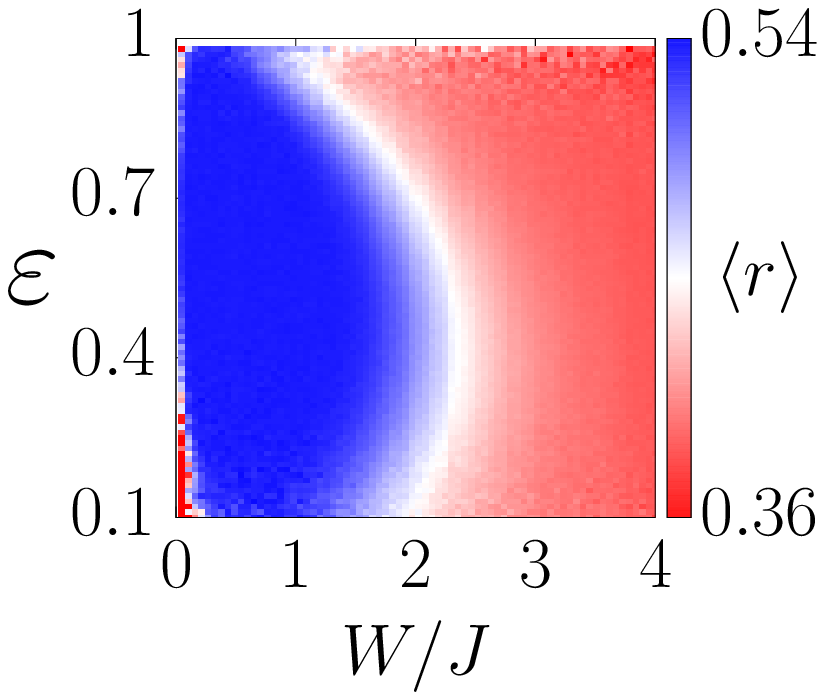}}

        \subfloat[][quinary]{\includegraphics[scale=0.6]{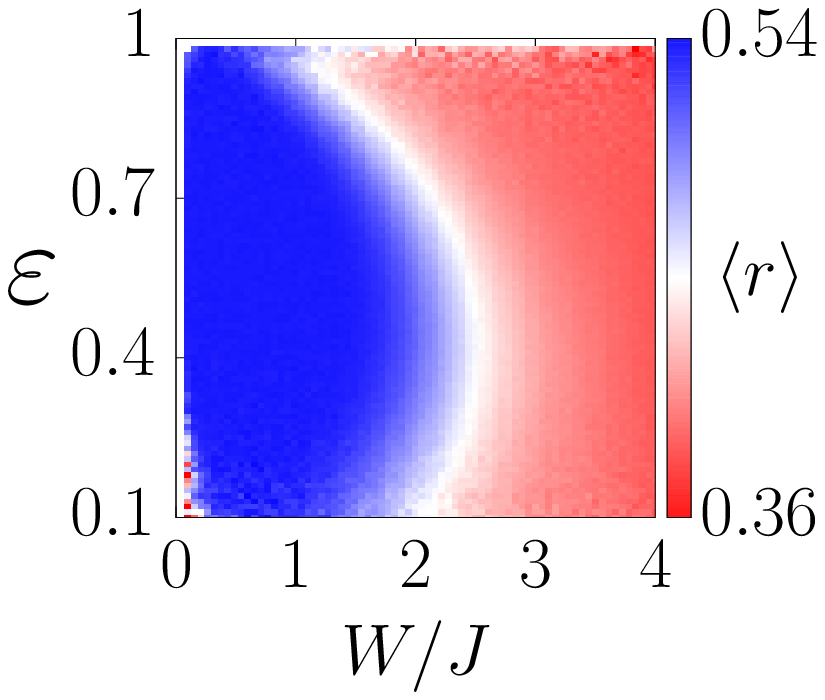}}\quad
        \subfloat[][uniform]{\includegraphics[scale=0.6]{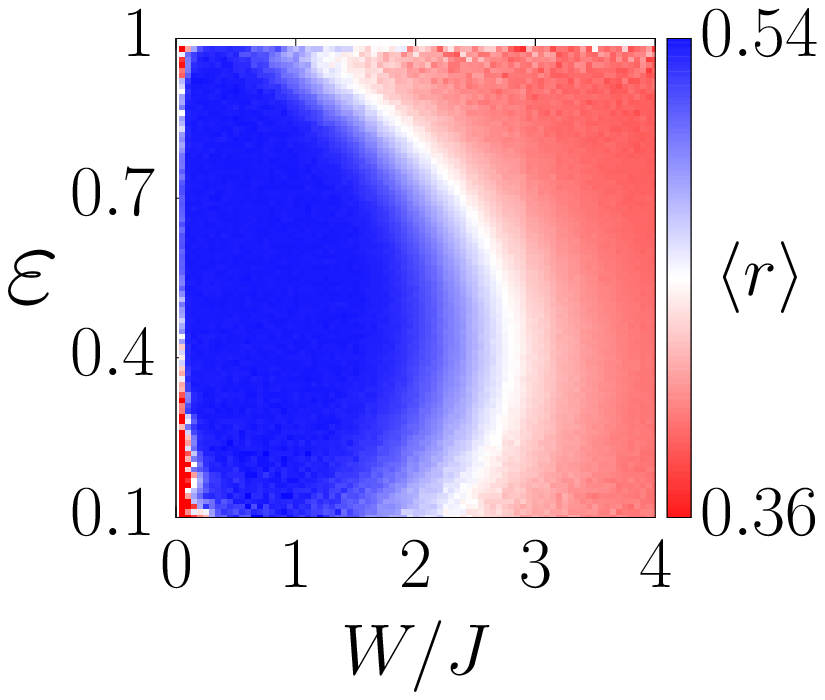}}

        \caption{Color plots of $\langle r\rangle$ in the $(\varepsilon, W/J)$ plane  for the various disorder types indicated in the figure. {Data are obtained for system size $L=16$ using periodic boundary conditions.} Blue regions correspond to the ergodic phase, red color indicates localized regions. The white color coincides with the intermediate value $\la r\ra = 0.45$.}\label{fig:r_bar_diagrams}
    \end{figure*}

Following \cite{Luitz15}, each spectrum is \emph{normalized} with transformation  $\varepsilon(E) = (E - E_{min}) / (E_{max} - E_{min})$. The spectrum is then  divided into 100 intervals $[\varepsilon_i-\delta\varepsilon, \varepsilon_i+\delta\varepsilon]$, with $\varepsilon_i = 0.01, 0.02, \ldots$ Next, we compute mean $\langle r\rangle$ values separately for all intervals in each spectrum.  Finally, values corresponding to the same intervals are averaged over 1000 realizations of disorder. As a result, we obtain diagrams $\langle r\rangle(\varepsilon, W)$ for the various disorder distributions studied. They are presented in \autoref{fig:r_bar_diagrams}.

The computed $\langle r\rangle$ values follow the GOE and Poisson predictions as limiting cases, blue regions indicate a delocalized phase where $\langle r\rangle \approx 0.53$, while red represents a localized phase with $\langle r\rangle \approx 0.39$. The careful reader will notice small regions for small disorder and low energy that unexpectedly show a localized behavior. This behavior may be traced back to the integrability of the Heisenberg model without  disorder. Also a similar behavior may be visible at the edge of the spectrum ($\varepsilon \approx 1$). Here the density of states is very low and the statistical significance of the corresponding data dubious.

The white color corresponds to the mean value of Poisson and GOE $\langle r\rangle$.  It gives an indication of the disorder value at which the transition occurs and reveals a significant dependence on energy (as in  \cite{Luitz15}). It is a hint of the existence of a localization edge. As we study  finite (small) systems sizes, the transition is necessarily smooth.
It is presently unknown if this cross-over evolves towards a true phase transition in the thermodynamic limit.

\begin{figure}
        \centering
        \includegraphics[width= 0.95\columnwidth]{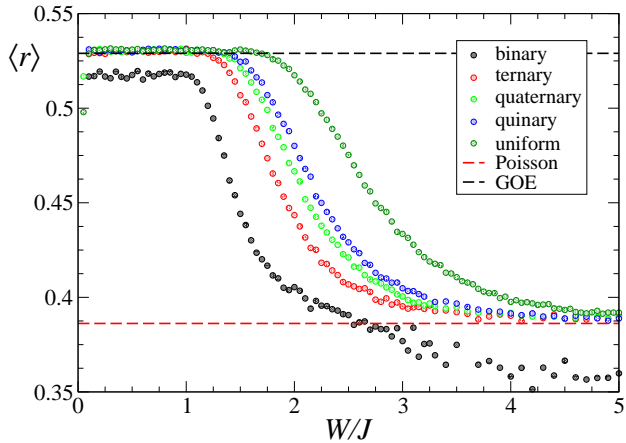}
        \caption{Mean ratio $\langle r \rangle$ {between consecutive energy level spacings} versus the disorder strength for a subset of data around $\varepsilon=0.5$. For all disorder distributions, except the binary one, a similar trend is observed: for low values of the disorder strength $W/J$, $\langle r\rangle$ takes the GOE value, for high $W/J,$ it follows the Poisson prediction. The data obtained for the binary distribution  deviate from that trend, the data do not reach GOE limit for small disorder while for strong disorder small ratios apparently push the mean below the value corresponding to the Poisson limit.}\label{fig:cut_r_bar}
    \end{figure}

It can be easily seen that diagrams differ by the size of the delocalized  {region}. With more possibilities of choice in the distribution, localization takes place for larger disorder strengths. Almost all distributions follow this behavior, although the binary distribution shows some deviation: the $\langle r\rangle$ values are smaller through the whole range of disorder strengths; at the beginning they are around 0.51, and for larger values of $W$ they go below the limiting Poisson value $\langle r\rangle \approx 0.39$. This can be better observed on a section of diagrams performed at value of normalized energy $\varepsilon = 0.5$ shown in \autoref{fig:cut_r_bar}.
Further examination of sections performed at different values of $\varepsilon$ show the same behavior.

\subsection{Binary disorder puzzle}
\label{sec:binary_puzzle}
The peculiar behavior observed for {the} binary distribution is in contrast {with} earlier studies (see Supplemental Material in \cite{Andraschko14} as well as results in \cite{Tang15}).
Those works reported results for the very same system studied with open boundary conditions (OBC) also for $L=16$. In particular, the GOE value of {$\langle r \rangle$} was observed for low disorder as well as an agreement with the Poisson prediction for strong disorder. {Both} these works took most energy levels into account (without the energy preselection) and, as it is apparent form Fig.~\ref{fig:r_bar_diagrams}, the average depends on energy in a nontrivial way. {Thus,} a direct comparison of our data with those of \cite{Andraschko14,Tang15} is not possible. Still the difference {between} their results {and} ours points towards different boundary conditions used: OBC were used in \cite{Andraschko14,Tang15} while we have used PBC which, typically, yield results closer to the thermodynamic limit, at least for the ground state of the system.

\begin{figure}
        \centering
        \includegraphics[width= 0.95\columnwidth]{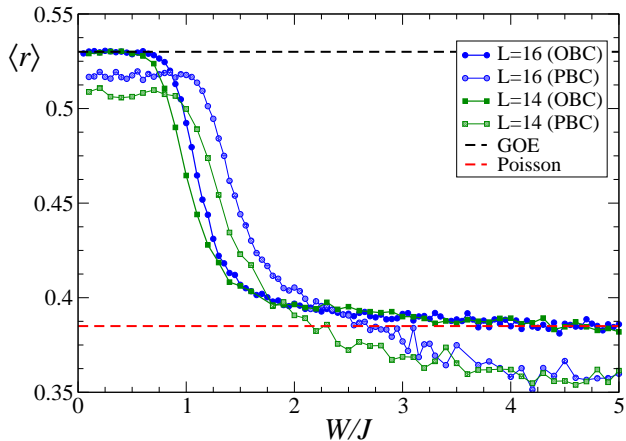}
        \caption{Mean ratio $\langle r \rangle$ {between consecutive energy level spacings} versus the disorder strength for a subset of data obtained for the Heisenberg chain with OBC around $\varepsilon=0.5$ for the binary disorder {(}$L=14$ and $L=16${)}. In contrast {with results displayed in Fig.~\ref{fig:cut_r_bar} for PBC, the mean  $\langle r \rangle$ value tends to the GOE prediction at low disorder strength $W/J$ and to the Poisson prediction at large disorder strength.}  \label{new1}}
\end{figure}

\begin{figure}
        \centering
        \includegraphics[width= 0.95\columnwidth]{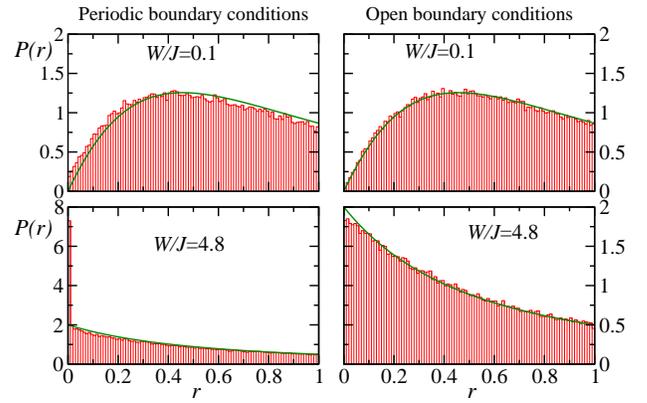}
        \caption{Histograms of {the} level spacing ratio $r$ for PBC (left column) and OBC (right column), for {the} Heisenberg chain {of length} $L=16${, in the presence of a binary disorder}. {The top}  row corresponds to small disorder ($W/J=0.1${)}
        while {the} bottom row {is} for a strong disorder $W/J=4.8$. Solid lines represent {the} GOE (for small disorder) and Poisson (large disorder) {distributions,} as given {in} \cite{Atas13}. Observe that while OBC lead to a good agreement with those distributions, for PBC we observe an excess of small spacing ratio, significant for small disorder yet quite spectacular ({especially in} the first bin) for large disorder. For explanation of this striking behavior, see \autoref{fig:split_symmetries} and the text.}\label{histo}
    \end{figure}

To check the influence of the boundary conditions, we have collected eigenvalues also for OBC - the results are shown in Fig.~\ref{new1} - again for energies around $\epsilon=0.5$.
Observe that both for $L=16$ and for smaller system {size} $L=14$, {the results} obtained {for} OBC agree with typical expectations {of the} GOE/Poisson {models}  for small/large disorder.  We have also {checked that our results averaged a broader range of energies are in excellent agreement with those of \cite{Tang15}} 
{Why is the mean ratio} so unexpectedly sensitive to the choice of boundary conditions? Some insight may be obtained looking at the full distributions of $r$, $P(r)$, instead of just the mean. Here analytic predictions for both limiting cases of GOE and Poisson statistics exist \cite{Atas13}. The distributions for exemplary small and large disorder amplitudes are shown in Fig.~\ref{histo}.
  
\begin{figure}
	\centering
	\includegraphics[width= 0.95\columnwidth]{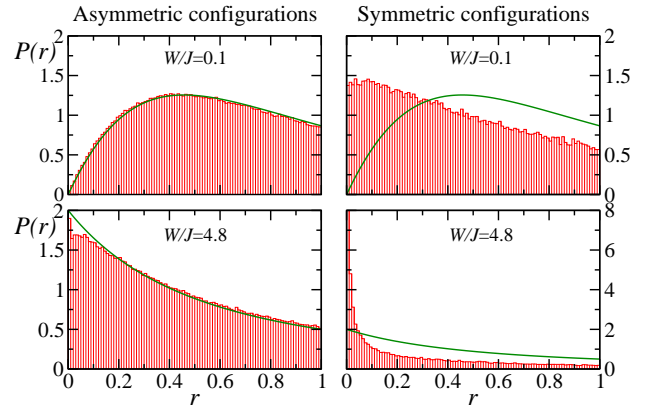}
	\caption{Histograms of the level spacing ratio $r$ for periodic boundary conditions, considering separately disorder configurations which are centro-symmetric around some point (right column) and asymmetric configurations (left column). For asymmetric configurations, the distributions are well predicted by the GOE model in the delocalized regime ($W/J=0.1$, upper plots)
	and the Poisson model in the localized regime ($W/J=4.8$, lower plots), shown by the solid green curves. In contrast, the symmetric disorder configurations are associated with completely different distributions, because the Hamiltonian can be block diagonalized, leading to a large excess of small spacings, thus small $r$ values. The relative abundance of symmetric configurations for periodic boundary conditions (12\% for $L=16$) is responsible for deviations observed in $\langle r\rangle$. The deviations are much smaller for ternary, quaternary,... distributions or for open boundary conditions. They eventually vanish in the thermodynamic limit. }\label{fig:split_symmetries}
\end{figure}

Clearly, PBC lead to an unexpected abundance of small spacing ratio, especially for large disorder. This surprising behavior may be understood realizing a peculiar feature of the binary disorder - the fact that just two disorder values are available. Suppose we have a symmetric disorder arrangement around a give{n} site. The parity operator around this center of symmetry commutes with the Hamiltonian, so that the latter can be block diagonalized in the symmetric and antisymmetric subspaces. In the delocalized regime, the energy spectrum in each subspace is described by a GOE model. Because the two subspaces are uncoupled, there is no level repulsion between a symmetric and an antisymmetric state, and one expects $P(r)$ not to vanish when $r\to 0.$  In order to test this hypothesis, we have specifically isolated the symmetric disorder configurations and numerically computed the corresponding $P(r)$ distributions. The results, shown in \autoref{fig:split_symmetries},
confirm that asymmetric disorder configurations are almost perfectly well described by the GOE prediction at low disorder, while symmetric configurations strongly deviate, with a non-zero probability of quasi-degeneracies. In the localized regime, there is additionally the situation where the localization center is further from the symmetry point than the localization length.
In such a case, the pair of symmetric/antisymmetric states will be quasi-degenerate, thus producing a peak around $r=0$ as observed in Figure~\ref{histo}. This is clearly demonstrated in
\autoref{fig:split_symmetries}, where asymmetric disorder configurations are well described by the Poisson distribution, while symmetric ones have a huge peak near $r=0.$ 
Such symmetric arrangements of disorder are quite abundant for a small size $L$. A simple analysis yields
$\approx 2L2^{L/2}$ {symmetric disorder arrangements} as compared to a total of $2^L$ configurations leading to about 12\% of excess small spacings for $L=16$.
For $L=16$ and periodic boundary conditions, there are 65536 binary disorder configurations, among which 5920 are centro-symmetric around some point. One must also add 1974 
configurations, where the disorder is antisymmetric with respect 
to some point; then, the product of the symmetry operator by spin reversal on each side ($S_i^z\to -S_i^z$) commutes with the Hamiltonian, again leading to a block diagonal structure for zero total magnetization. 

For OBC, symmetric configurations exist only if
 		the symmetry point is  at the middle of the chain,  so it is $L$ times less abundant; the effect on $\langle r \rangle$ is much smaller, of the order of the statistical error bars. Similarly, one may check that the effect is much smaller for ternary, or higher valued disorder. The excess of quasi-degeneracies is clearly an unexpected and surprising  
artifact of the PBC for binary disorder. {Thus,}
we {will} use in the following OBC for the binary disorder data.  

 There are clever ways to avoid the spurious effect of symmetric configurations, by modifying the boundary conditions (see Supplementary Material in \cite{Mondaini15}) Since our aim is primarily aimed at locating the region of transition between extended and localized states for different disorders rather than a detailed comparison of level statistics, we refrained from modifying the boundary conditions.

\subsection{Variance renormalization}
\begin{figure*}
        \centering
        \subfloat[][binary]{\includegraphics[scale=0.6]{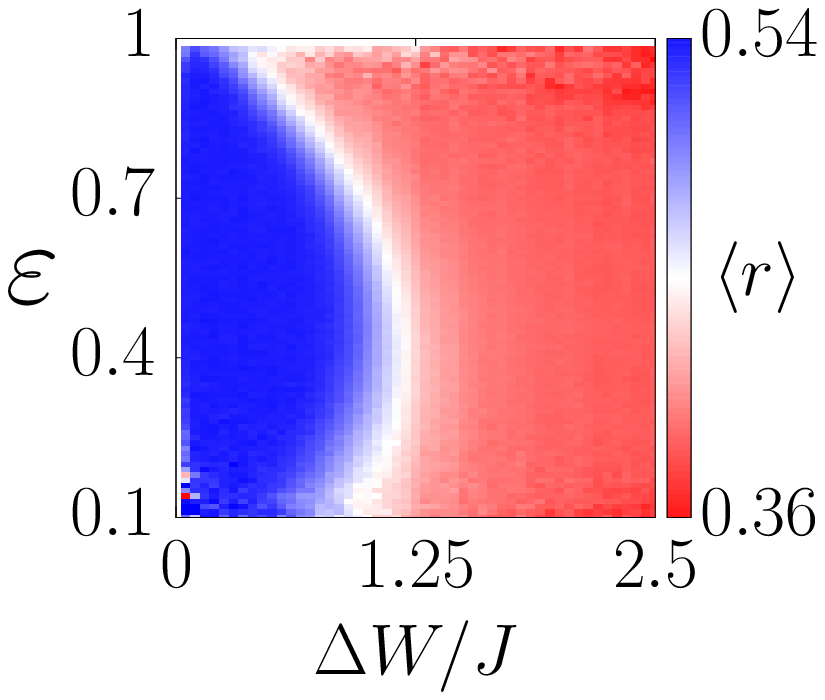}}\quad
        \subfloat[][ternary]{\includegraphics[scale=0.6]{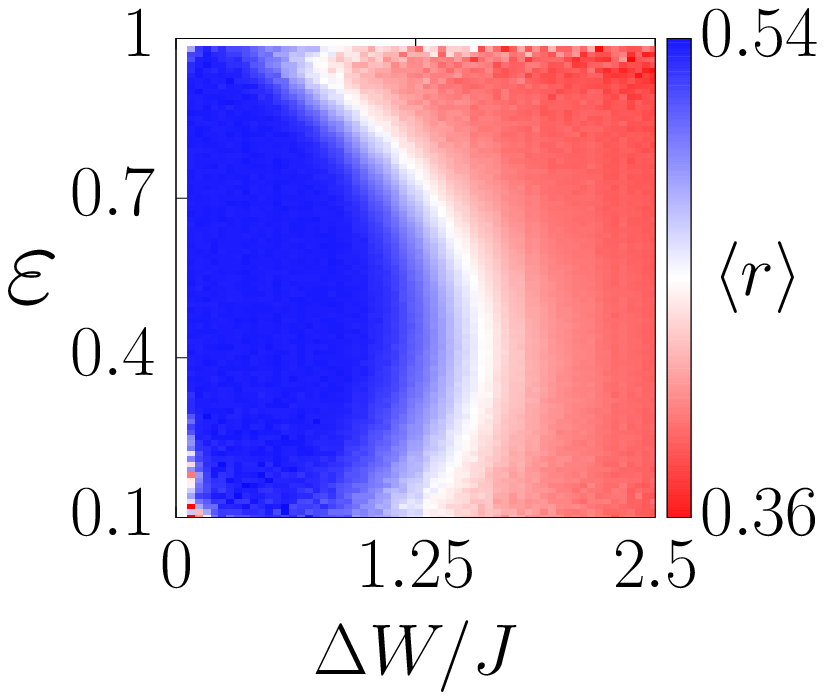}}\quad
        \subfloat[][quaternary]{\includegraphics[scale=0.6]{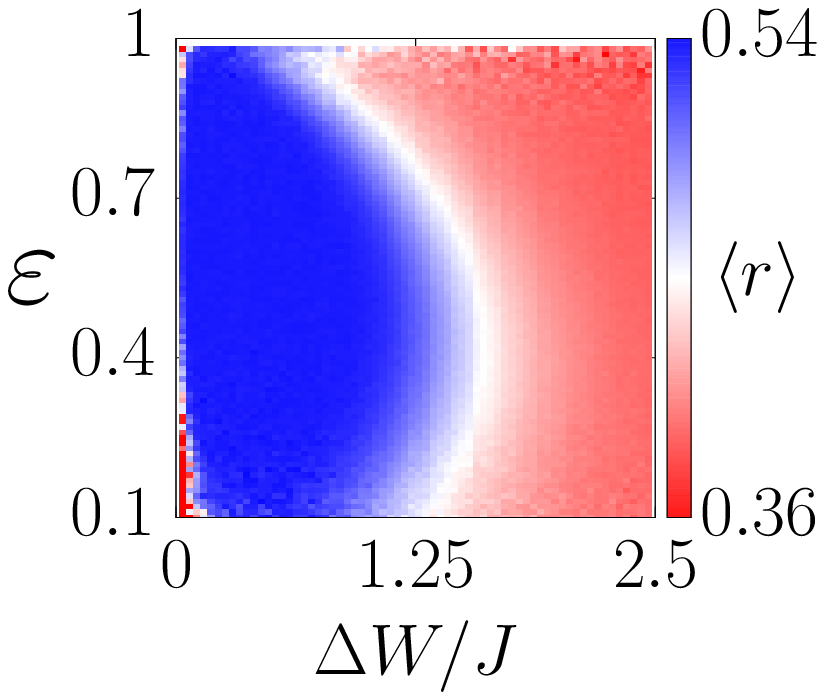}}

        \subfloat[][quinary]{\includegraphics[scale=0.6]{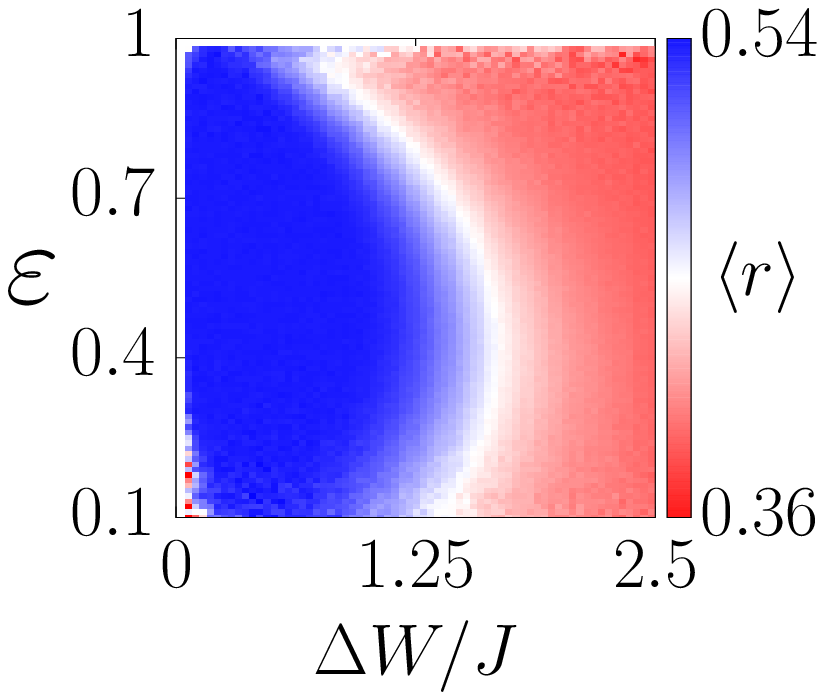}}\quad
        \subfloat[][uniform]{\includegraphics[scale=0.6]{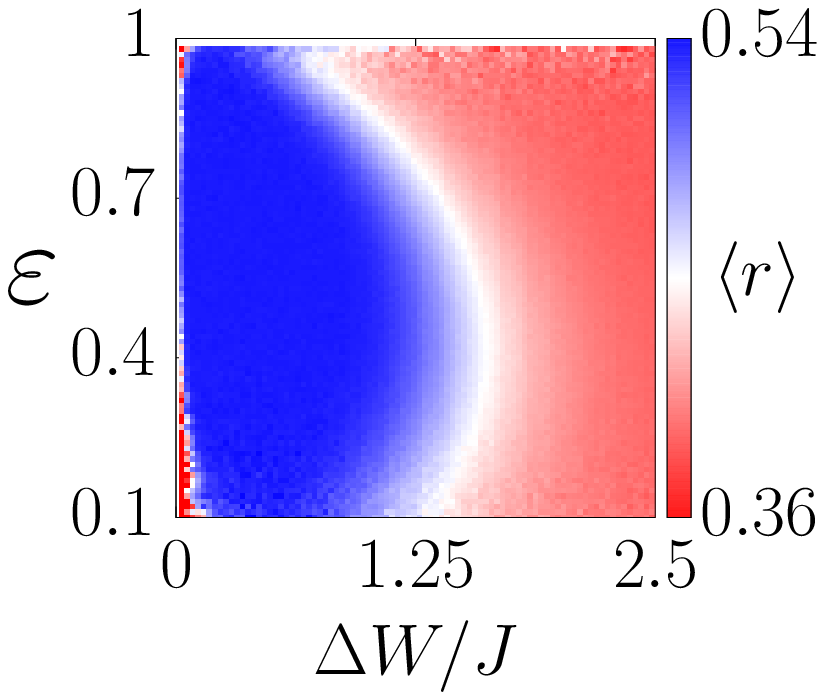}}

        \caption{Same as Fig.~\ref{fig:r_bar_diagrams} but after rescaling with respect to the variance of different distributions. {For the binary distribution (a), we use the results obtained for open boundary conditions as discussed in the text, while we keep periodic boundary conditions for other distributions (b-e).} Observe that all the plots look very similar. A careful observer will notice that the lobe is  smaller for the binary disorder.}\label{fig:rescaled_r_bar_diagrams}
    \end{figure*}

After overcoming this artifact of the binary disorder, we observe that the transition to the localized behavior occurs for larger $W/J$ for larger disorder multiplicity. This observation is crucial for the next step of our analysis. Distributions have zero mean, but have different values of the variance. For Anderson localization in a weak disorder (Born approximation), the localization length depends only on the potential average and its variance. Following this, we compare results at points of equal variance, i.e. different disorder strength. It is sufficient to perform a simple rescaling by the standard deviation of the distribution:
For a $N$-ary disorder with $N$ values equally spaced in the $[-W,W]$ interval, it is:
    \begin{equation}
        \Delta W = \sqrt{\langle w_i^2\rangle} = W \sqrt{\frac{N+1}{3(N-1)}},
    \end{equation}
which leads to:
    \begin{center}
        \begin{tabular}{r | l}
            binary & $W \rightarrow \Delta W = W$ \\ \hline
            ternary & $W\rightarrow \Delta W = \sqrt{2/3} \, W $ \\ \hline
            quaternary & $W\rightarrow \Delta W = \sqrt{5/9}\, W$ \\ \hline
            quinary & $W\rightarrow \Delta W = \sqrt{1/2}\, W$ \\ \hline
            uniform $N\to \infty$ & $W\rightarrow \Delta W = \sqrt{1/3}\, W$
        \end{tabular}
    \end{center}
This transformation allows us to plot  maps similar to \autoref{fig:r_bar_diagrams}, but in the space of the normalized energy and the standard deviation of the disorder strength. They are presented in \autoref{fig:rescaled_r_bar_diagrams} {(note that, following the discussion in section \ref{sec:binary_puzzle}, we use OBC for the binary disorder and PBC for the other disorders)} From ternary disorder upwards, the models behave almost identically. It suggests that the transition from the ergodic phase to the MBL one is identical to the one observed for uniform disorder. The lobe for the binary disorder looks also similar, being slightly smaller, with the border shifted to smaller variance values.

\begin{figure}[!h]
        \centering
        \includegraphics[width= 0.95\columnwidth]{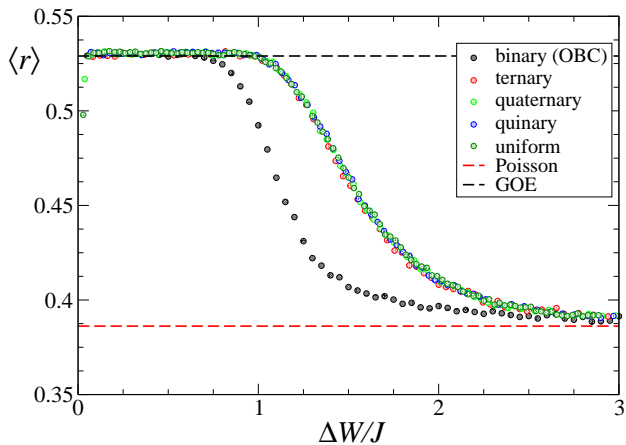}
        \caption{Mean ratio $\langle r \rangle$ as a function of the standard deviation of the disorder strength for each distribution, for $\varepsilon=0.5$. Such a rescaling reveals practically identical curves for all disorder types, except for binary disorder which shows a significant deviation. The statistical errors of the data points are comparable to the symbol
        size.
      }
         \label{fig:rescaled_cut_r_bar}
    \end{figure}

To show the quality of {the} agreement between the various distributions, we show in  \autoref{fig:rescaled_cut_r_bar} a detailed comparison for $\varepsilon = 0.5$. 
    \begin{figure}[!h]
        \centering
        \includegraphics[width= 0.95\columnwidth]{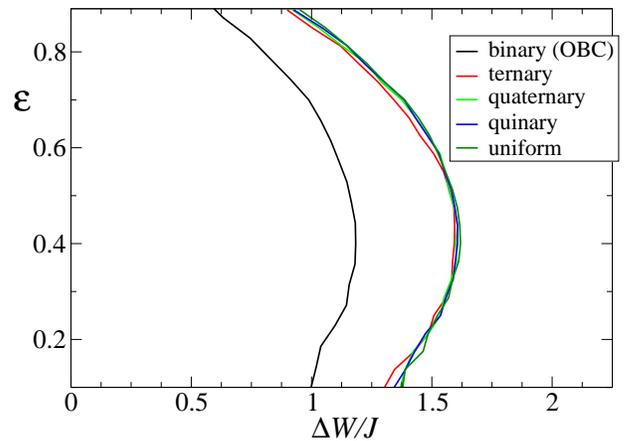}
        
        \caption{Values of the rescaled energy corresponding to the mean ratio $\langle r \rangle$=0.45 (i.e. in the middle of the ergodic-MBL transition)  versus standard deviation of the disorder strength for different distributions. Again, for all distributions, except the binary one, the transition almost coincides with the one for the uniform distribution.}\label{fig:mean_ratio_eps}
    \end{figure}
This confirms that the ternary, quaternary and quinary distributions yield similar predictions to the random uniform continuous distribution, after renormalization to the same value of the standard deviation. Similar coalescence of curves after rescaling of variance is observed at other $\varepsilon$ values. This is visualized further in \autoref{fig:mean_ratio_eps} where a cut at $\langle r\rangle=0.45$ for different $\varepsilon$ values is presented. The errors for the data for small $\epsilon$ values, due to smaller density of states in this region,  are necessarily bigger than those for $\epsilon \approx 0.5$. 

To conclude this part of the paper, we want to stress out that, if one is interested in the statistical properties of energy levels, it suffices to use ternary disorder in place of the continuous distribution. In such a case, instead of using the disorder strength as a parameter of a model, one should use the standard deviation. Additionally, our analysis shows that the binary distribution differs significantly from other types of discrete distributions. This is important in view of the attempted quantum averaging with {an} ancillary spin {one-}half  \cite{Enss17}. Our results strongly indicate that unit spins (leading to ternary disorder) should be used instead if one wants to draw some conclusions for the continuous random uniform distribution.

\section{Time evolution of the order parameter}

Consider now the temporal evolution of the spin chain prepared initially in the N\'eel state $\ket{\psi(0)} = \ket{\uparrow\downarrow\uparrow\downarrow\ldots}$.  As the order parameter, we choose  the imbalance $I$ (i.e. twice the staggered magnetization of the sample) defined as:
    \begin{equation}
        I(t) = \frac{4}{L} \sum_{i=1}^L \bra{\psi(0)}S^z_{i}(0) S^z_{i}(t)\ket{\psi(0)},
        \label{eq:def_imbalance}
    \end{equation}
which coincides with the observable widely used in experiments \cite{Schreiber15}. Initially $I(0) = 1$. In the long time limit, it  should vanish in the ergodic situation due to the ETH, while it may remain non-zero for localized systems. This \emph{final imbalance} value depends on the type of the disorder used as well as its strength, it is defined as a long time average:
    \begin{equation}
	\overline{I} = \lim_{t\rightarrow\infty}\frac{1}{t}\int_{t'=0}^t I(t')\text{d}t'.
    \end{equation}

    \begin{figure}
        \centering
        \includegraphics[width= 0.95\columnwidth]{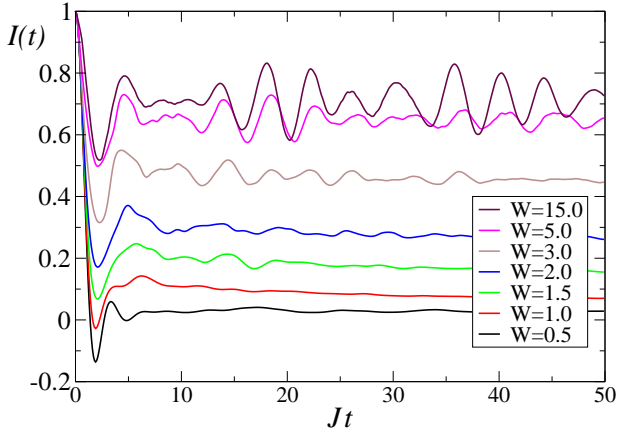}
        \caption{Examples of the imbalance $I(t)$, Eq.~(\ref{eq:def_imbalance}) for quaternary disorder, and various disorder strengths.}\label{fig:magnetization_examples}
    \end{figure}

    \begin{figure}
        \centering
        \includegraphics[width= 0.95\columnwidth]{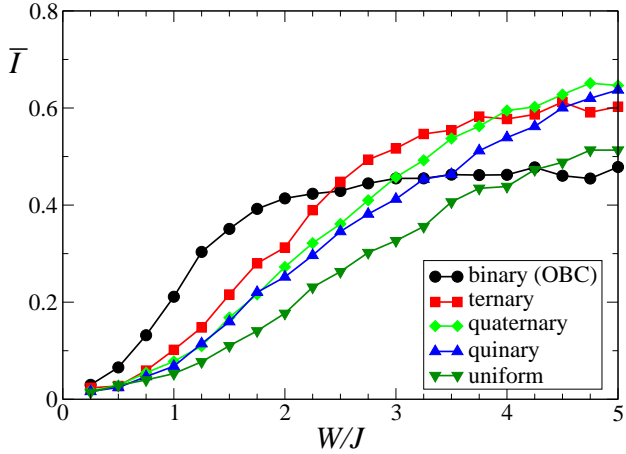}
        \caption{Final imbalance as a function of the disorder strength for different types of disorder indicated in the figure.}\label{fig:final_magnetization}
    \end{figure}

Observe that this experimentally based procedure may be replaced in our case by an exact long time average by considering only diagonal contributions to \eqref{eq:def_imbalance}. Still to simulate the experiment we have performed
simulations up to time $t=50$. The results are averaged over 250 realizations of disorder for every distribution. Exemplary results of $I(t)$ for the quaternary disorder distribution are presented in \autoref{fig:magnetization_examples}. After an initial drop and some transients, the imbalance reaches a stationary value, with oscillations around it, in particular for strong disorder. Non vanishing values of the imbalance at small $W/J$ are caused by finite size effects. To compute the value of final imbalance, we averaged  the data for $t\in [20,50]$. \autoref{fig:final_magnetization} shows the comparison of the final  imbalance vs. the disorder strength, for various types of disorder.  All disorder models show a similar behavior: for large $W$, the imbalance does not vanish at long time. However, the saturation takes place faster for distributions with lower number of possible choices. Similarly, the maximal value of the final imbalance is smaller for distributions with a larger 
variance.

    \begin{figure}
        \centering
        \includegraphics[width= 0.95\columnwidth]{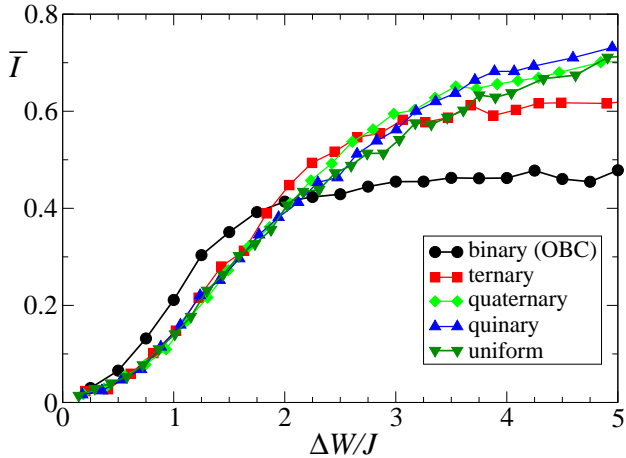}
        \caption{Final imbalance as a function of the disorder standard deviation.}
        \label{fig:rescaled_final_magnetization}
    \end{figure}

These dependencies suggest that, again, the proper measure to characterize the strength of the disorder is using the standard deviation of its distribution. Indeed a comparison of models as a function of $\Delta W/J$ shown in \autoref{fig:rescaled_final_magnetization}, reveals that the final imbalance behaves in a similar way in the transient regime. The limiting imbalance at large $\Delta W/J$ depends on the disorder distribution, and deviates from the universal (or limiting) curve of the continuous distribution.

\noindent{\textit{Analytic estimation of the final imbalance at large disorder.}}
In the regime of high disorder strengths, strong localization with a short localization length is expected. Thus the strongest impact on the dynamics of the imbalance should come from the coupling between nearest sites. The simplest model involves two  neighboring sites of the chain only. Such a model was successfully applied for studies of MBL with randomly interacting bosons \cite{Sierant17,Sierant17b}.  We apply the very same idea to our spin system.

Consider the 2-sites N\'eel state basis $\ket{\downarrow\uparrow}$, $\ket{\uparrow \downarrow}$ where the Hamiltonian for 2 adjacent spins takes form:
    \begin{equation}
        \mathcal{H}' = \frac{1}{2}
        \begin{pmatrix}
            w_2-w_1 - J/2 & J\\
            J & w_1-w_2 - J/2
        \end{pmatrix},
\end{equation}
where $w_i$ are the disorder values at the two sites. As the initial state, we take $\ket{\psi(0)} = \ket{\downarrow\uparrow}$. A simple calculation shows that the imbalance is given by:
    \begin{equation}
        I(t) = \frac{\Delta ^2+ \cos \left(J t \sqrt{\Delta ^2+1}\right)}{\Delta ^2+1},
    \end{equation}
where $\Delta = (w_2 - w_1)/J$. Then, the final imbalance is simply calculated as the long time average:
    \begin{equation}
        \overline{I}= \lim_{t\rightarrow \infty}\frac{1}{t}\int_{t'=0}^t I(t') \text{d}t' = \frac{\Delta ^2}{ \Delta ^2+1}.
        \label{eq:imbalance_2site_delta}
    \end{equation}
Having this closed form expression, it is straightforward to calculate averaged values of $\overline{I}$ for our disorder models. Summing all possible values of $\Delta$ for a $N$-ary disorder final imbalance takes {the} following form:
    \begin{equation}
        \overline{I} = \frac{2}{N^2} \sum_{k=1}^{N-1}{(N-k)\frac{\left(\frac{2k(W/J)}{N-1}\right)^2}{1+\left(\frac{2k(W/J)}{N-1}\right)^2}}
        \label{eq:sum}
    \end{equation}
 In the limit of large $W/J$, we can perform an asymptotic expansion in powers of $J/W:$
    \begin{equation}
    \overline{I} = \frac{N-1}{N}- \frac{(N-1)^2}{2N^2(W/J)^2}\sum_{k=1}^{N-1}\frac{N-k}{k^2}.\label{II}
    \end{equation}
This gives:
    \begin{equation}
    \begin{aligned}
    &\text{binary: } & \overline{I} &= \frac{1}{2}  - \frac{J^2}{8 W^2}  + {\mathcal O}\left({\frac{J^4}{W^4}}\right), \\
    &\text{ternary: } & \overline{I} &= \frac{2}{3} - \frac{J^2}{2 W^2} +  {\mathcal O}\left({\frac{J^4}{W^4}}\right),\\
    &\text{quaternary: } & \overline{I} &= \frac{3}{4} - \frac{65 J^2}{64 W^2} +  {\mathcal O}\left({\frac{J^4}{W^4}}\right),\\
    &\text{quinary: } &\overline{I} &= \frac{4}{5} - \frac{29 J^2}{18 W^2} +{\mathcal O}\left({\frac{J^4}{W^4}}\right).
    \end{aligned}\label{eq:final_analytical}
    \end{equation}
\autoref{II} is limited to situations where $W/J\gg N$. It is interesting to look at the large $N$ limit assuming that $1 \ll W/J \ll N$. In such a case it is possible to replace the discrete sum in \autoref{eq:sum} by a continuous integral:
\begin{equation}
    \overline{I} = 1 - \frac{1}{N} - \frac{2}{N^2}\int_0^N (N-x)\frac{(N-1)^2}{(N-1)^2 + (2 x w)^2} \text{d}x.
\end{equation}
Because only terms up to $x=N/(W/J)$ contribute to the integrand, $(N-x)$ can be replaced by $N$. Extending the upper integral limit to infinity one gets as a result:
\begin{equation}
    \overline{I} = \lim_{N\rightarrow \infty}\frac{N-1}{N}\left(1 - \frac{\pi J}{2 W}\right) = 1 - \frac{\pi J}{2 W}
\end{equation}
    \begin{figure}
        \centering
        \includegraphics[width= 0.95\columnwidth]{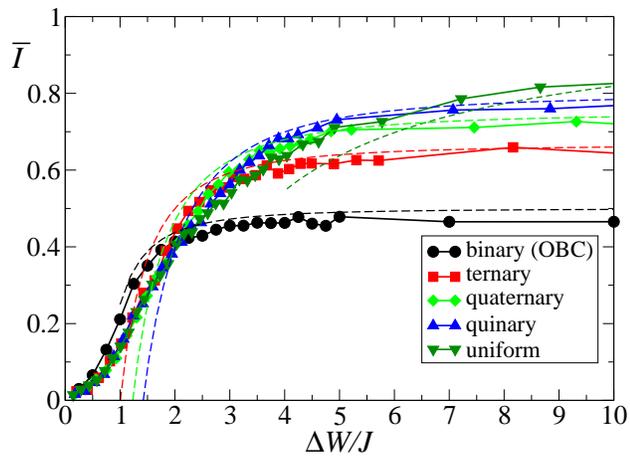}
        \caption{Comparison of numerical (solid lines) and theoretical (dashed lines) predictions for the 2-site model (\autoref{eq:final_analytical}).} \label{fig:final_magnetization_model}
    \end{figure}
recovering the leading term of the analytic result obtained for uniform distribution (\autoref{eq:imbalance_2site_delta} ):
    \begin{equation}
\begin{aligned}
     \overline{I} = 1 + \frac{\log(4(W/J)^2 + 1)}{4 (W/J)^2} - \frac{\arctan(2 W/J)}{W/J} \\
      \approx 1 - \frac{\pi J}{2 W} + \mathcal{O}\left(\frac{J^2}{W^2}\right).
\end{aligned}
    \end{equation}

Before comparing our prediction with numerical results, it is important to notice {that} we considered systems having only 2 sites. In a spin chain, each spin has 2 neighbors, and hopping may occur to the left or to the right. The effect doubles the probability of losing imbalance. It can be included in \autoref{eq:final_analytical} multiplying by 2 the second factor for each disorder type (the same correction was  used for the similar model for bosons \cite{Sierant17,Sierant17b}). Such a correction allows us to compare  numerical results with the asymptotic predictions {in} \autoref{fig:final_magnetization_model}. The agreement is good, especially for large values of $\Delta W/J$, in line with the assumptions of our model.

\section{Conclusions}
We have studied different discrete disorder models and their effect on the MBL phenomenon via an analysis of level statistics and of the long time dynamics.
For level spacings ratios, we have observed, upon increasing the disorder strength, the transition from extended (ergodic) states (with GOE statistics) to localized states (with Poisson statistics). By appropriately rescaling the disorder strength by its standard deviation, we have observed that all $N$-ary distributions with $N>2$ lead to the same behavior than the random uniform distribution.
The binary disorder distribution has a slightly different behavior. On the one hand, level statistics for small systems are
more sensitive to the boundary conditions. On the other hand, even when this effect is eliminated by using OBC, there are still
quantitative differences. This is important in the context of quantum disorder simulation protocol \cite{Paredes05,Enss17} which commonly uses ancillary 1/2-spins leading to binary disorder. Our findings indicate that, in order to simulate the effects of a random uniform distribution, ternary or higher $N$-ary disorder models should be used (corresponding to $S\!=\!1$ or larger for the ancillary spins in the quantum averaging procedure).

Interestingly, the time evolution of the order parameter (imbalance) shows a slightly different behavior. At strong disorder, the imbalance saturates at a value depending on the disorder model. This saturating limit is well grasped by a simple two site theory. This model also describes the final imbalance  in the strong disorder limit.
For smaller disorder values, the rescaling of the disorder strength again produces a universal dependence of the imbalance.

\begin{acknowledgments}
This work was supported by Polish National Science Centre via project No. 2015/19/B/ST2/01028. Support by PL-Grid Infrastructure and by EU via project QUIC (H2020-FETPROACT-2014 No. 641122) is  also acknowledged.
\end{acknowledgments}

%

\end{document}